\documentclass{ws-ijmpb}

\usepackage{amsmath}
\usepackage{amssymb}
\usepackage{graphics}

\begin{document}

%
\catchline{}{}{}{}{}
%

\title{Quantum corrections to the dynamics of the Bose-Einstein condensate \\
in a double-well potential} 
\author{Yan~Xu$^{1,2}$, Wei~Fan$^{1,\ast}$ and Bing~Chen$^{1}$}
\address{$^{1}$College of Science, Shandong University of Science and Technology, \\
579 Qianwangang Road, Economic and Technical Development Zone, \\
Qingdao, 266510, China}
\address{$^2$ Center for Quantum Technologies, National University of Singapore,\\
2 Science Drive 3,Singapore 117542, Republic of Singapore\\
$^\ast$ashitakatosan@gmail.com
}

\maketitle

\begin{abstract}
        The dynamics of the Bose-Einstein condensate (BEC) in a double-well 
        potential is often investigated under the mean-field theory (MFT).  
         This works successfully for large particle numbers with dynamical stability. 
         But for dynamical instabilities, quantum corrections to the 
         MFT becomes important [Phys.Rev.A \textbf{64}, 013605(2001)]. 
        Recently the adiabatic dynamics of the double-well BEC is investigated 
         under the MFT in terms of a dark variable [Phys.Rev.A \textbf{81}, 043621(2010)],
        which generalizes the adiabatic passage 
        techniques in quantum optics to the nonlinear matter-wave case.  We 
        give a fully quantized version of it using second-quantization 
        and introduce new correction terms from higher order interactions 
        beyond the on-site interaction, which are interactions between the
        tunneling particle and the particle in the well and interactions between the
        tunneling particles. 
        If only the on-site interaction is considered , this reduces 
        to the usual two-mode BEC. 
\end{abstract}


\maketitle

\section{Introduction}

The dynamics of the Bose-Einstein condensate (BEC) is often investigated under 
the mean-field theory (MFT), which provides a classical field equation for the 
nonlinear matter wave and is the classical limit of the quantum theory in 
large number limit.  MFT works successfully in predicting experimental results 
as the particle numbers of BEC is very large, so the classical limit captures 
the essence. In the vicinity of a mean-field dynamical instability, however, the 
quantum correction to the MFT becomes important~\cite{1}. They 
provide accurate predictions for the dynamics by combining the mean-field with 
the fluctuations. As most investigations are done under the MFT, it is 
interesting to look at their behavior under quantum corrections.    

On the other hand, the analogy to quantum optics is noticed in investigating BEC, 
because of its macrosscopic coherence that allowes us to view it as a large atom.
A recent example 
is the discovery of a dark variable for the double-well BEC by
Ottaviani et al~\cite{2}, which is a nonlinear matter-wave extension 
of Vitanov and Shore's
work~\cite{3} of a two-state atom, where a dark variable is found by
looking into the analogy between the optical two-state Bloch equation and
the three-state stimulated Raman adiabatic passage (STRAP) equations~\cite{4}.
They compose three
variables analogous to the Bloch spin from the ground states of the two wells and the
dark variable is found by comparing the equation of the three variables with the STRAP
equation. The adiabatic splitting, transport and self-trapping of the double-well BEC are
investigated using this dark variable~\cite{2} under the MFT.
The STRAP can be utilized to transport BEC in multiple 
wells~\cite{5}$^-$\cite{9}.

In the paper, we give a fully quantized version of the work of Ottaviani et 
al.  We introduce further quantum corrections by including higher order 
interactions beyond the on-site interaction.  If these quantum corrections are 
turned off, this model would reduce to the two-mode BEC widely used in 
quantum entanglement~\cite{10}$^-$\cite{19}.

\section{The on-site Interaction}

In MFT,  the two-mode approximation is applied to the GP
equation and the dynamics of the double-well BEC~\cite{2} 
is 
\begin{equation}
\mathcal{H} \left( \begin{array}{c} c_L \\
c_R \end{array} \right) =  i\frac{d}{dt} \left( \begin{array}{c} c_L\\
c_R \end{array} \right),
\end{equation}
where the Hamiltonian is given by
\begin{equation}
\mathcal{H} = \left( \begin{array}{cc} \epsilon_L + U_L \left\vert c_L \right\vert ^{2} & \Omega\\
\Omega & \epsilon_R + U_R \left\vert c_R \right\vert ^{2} \end{array} \right).
\end{equation} 
Introducing the field operator $\hat{\Psi}\left(\vec{r}, t\right)$, 
 the Hamiltonian can be written as 
 \begin{eqnarray}
 \mathcal{H} & = & \mathcal{H}_0 + \mathcal{H}_{int} \nonumber\\
                        & =   &{\displaystyle \int d^{3}r\, \hat{\Psi}^{\dagger} H_0 \hat{\Psi}} \nonumber\\
                        &    &{}+ \frac{g}{2} {\displaystyle \int d^{3}r\, \hat{\Psi}^{\dagger} \hat{\Psi}^{\dagger} \hat{\Psi} \hat{\Psi}},
 \end{eqnarray}
where \mbox{$H_0 = - \nabla^{2} / 2m + V\left(\vec{r}, t\right)$} is 
 the single particle Hamiltonian with $V\left(\vec{r},  t\right)$ the 
 double-well potential and \mbox{$g = 4 \pi a_{s} / m $}  is the 
nonlinear interaction parameter with $a_{s}$ the $s$-wave scattering 
length.
Only spatial degrees of
freedom and two-body interactions are considered.  
Under the two mode approximation, where all  modes are omitted except the condensate 
modes, the field operator can be expanded  
\mbox{$\hat{\Psi}\left(\vec{r},t\right) = \hat{a}_L\left(t\right) \phi_L\left(\vec{r}\right) + \hat{a}_R\left(t\right)
\phi_R\left(\vec{r}\right)$}, where $\hat{a}_{L, R}$ and $\phi_{L, R}$ are the annihilation operator and the 
ground state of the left and the right well respectively. The overlap between the two 
ground states is  neglected because it is small 
compared with the on-site part, 
$\phi_L^{\ast} \phi_R \ll \left\vert \phi_L\right\vert^2$. 
Now the second quantized Hamiltonian becomes 
\begin{eqnarray} \label{hamiltonianBoson}
\mathcal{H} & = & \epsilon_L \hat{a}_L^{\dagger} \hat{a}_L  + \Omega \hat{a}_L^{\dagger} \hat{a}_R +
\frac{U_0}{2} \hat{a}_L^{\dagger} \hat{a}_L^{\dagger} \hat{a}_L \hat{a}_L\nonumber\\
                      &     & {} + \epsilon_R \hat{a}_R^{\dagger} \hat{a}_R  + \Omega \hat{a}_R^{\dagger}
                      \hat{a}_L + \frac{U_0}{2} \hat{a}_R^{\dagger} \hat{a}_R^{\dagger} \hat{a}_R \hat{a}_R,
\end{eqnarray}
with the parameters 
\begin{eqnarray}
\epsilon_{L, R} = {\displaystyle \int d^{3}r\, \phi_{L, R}^{\ast} H_0 \phi_{L, R}}, \\
\Omega = {\displaystyle \int d^{3}r\, \phi_L^{\ast} H_0 \phi_R} = {\displaystyle \int d^{3}r\, \phi_R^{\ast} H_0 \phi_L}, \\
U_0 = g {\displaystyle \int d^{3}r\, \left\vert \phi_L \right\vert^{4}} = {\displaystyle \int d^{3}r\, \left\vert \phi_R \right\vert^{4}}.
\end{eqnarray}
with $\epsilon_{L, R}$ the chemical potential of the two wells respectively 
and $U_0$ the nonlinear on-site interaction between particles in the same well.  
The tunneling rate $\Omega$ between the two wells is negtive, but it can also 
be defined to be positive by adding corresponding minus signs in the 
Hamiltonian.  

It is convenient to rewrite the Hamiltonian by the Schwinger angular momentum operators 
$\hat{J}_x = \left( \hat{a}_R^{\dagger} \hat{a}_L + \hat{a}_L^{\dagger} \hat{a}_R \right) / 2,
\hat{J}_y = \left( \hat{a}_R^{\dagger} \hat{a}_L - \hat{a}_L^{\dagger} \hat{a}_R \right) / 2 i,
\hat{J}_z = \left( \hat{a}_R^{\dagger} \hat{a}_R - \hat{a}_L^{\dagger} \hat{a}_L \right)/2,$
where $\hat{J}_{x, y}$ corresponds to the correlation between the 
two wells and $\hat{J}_z$ is the particle number
difference between the two wells. The Hamiltonian under these angular momentum 
operators becomes
\begin{eqnarray}\label{simpleAngular}
\mathcal{H} & = & \frac{E}{2} N + \epsilon \hat{J}_{z} + 2 \Omega \hat{J}_{x} + U_0 \left( \hat{J}_z^2 + \frac{N^2}{4} -
  \frac{N}{2} \right) \nonumber\\
                      & = & U_0 \hat{J}_z^2 + \epsilon \hat{J}_z + 2 \Omega \hat{J}_x,
\end{eqnarray}
with $N = N_L + N_R $ the total particle number, $E = \epsilon_L + \epsilon_R $ 
the sum of the two chemical potentials and 
$ \epsilon =\epsilon_R - \epsilon_L$ the difference of the two chemical potential.
Terms containing $N$ and $E$ are assumed to be conserved and are neglected 
from the Hamiltonian.   
This is the often used Hamiltonian in quantum entanglement, which only the on-site interaction is included.
The dynamics of the system manifests in the evolution of these angular momentum operators 
\begin{equation} \label{simpleDynamics}
\frac{d}{dt} \left( \begin{array}{c} \hat{J}_x\\
\hat{J}_y\\
\hat{J}_z \end{array} \right) = \left( \begin{array}{ccc} - i U_0 & - \left( \epsilon + 2 U_0 \hat{J}_z \right) & 0\\
\epsilon + 2 U_0 \hat{J}_z & - i U_0 & - 2 \Omega\\
0 & 2 \Omega & 0 \end{array} \right) \left( \begin{array}{c} \hat{J}_x\\
\hat{J}_y\\
\hat{J}_z \end{array} \right).
\end{equation}
Equation ``Eq.~(\ref{simpleDynamics})'' is the second quantized version of the dynamics
given in Ref.~\refcite{2}, where it is described by the 
evolution equation of the
Bloch spin vector~\cite{20},
\begin{equation}\label{firstQuantized}
\frac{d}{dt} \left(\begin{array}{c} \mu\\
\nu\\
\omega \end{array} \right) = \left( \begin{array}{ccc} 0 & -(\epsilon + U\omega) & 0\\
(\epsilon + U\omega) & 0 & -2\omega\\
0 & 2\omega & 0 \end{array} \right) \left(\begin{array}{c} \mu\\
\nu\\
\omega \end{array} \right),
\end{equation}
The elements of the Block spin vector $\mu, \nu \mbox{and } \omega$  are composed from
the two modal populations. 

The two evolution equations would resemble each other with
 the angular momentum
operator corresponding to the Bloch spin vector,
if we normalize the angular
momentum operator by $N / 2$, 
 $ 2 \hat{J}_{x, y, z} / N \thicksim u, v, w $. The nonlinear interaction
parameter becomes  $ g = 4 \pi N a_s / 2 m$ under this normalization, which 
becomes the same $g$  in ``Eq.~(\ref{firstQuantized})''.
Without the normalization, the adiabatic splitting, transport and 
self-trapping can be given by the motion of the angular momentum operator
on the Bloch sphere with radius $j = N / 2$,   while in
Ref.~\refcite{2} it corresponds to the motion of the Bloch spine in
the unit sphere. The adiabatic transport corresponds to the variance of $J_z$ from
$- N / 2$ to $0$ then to $N / 2$ and the self-trapping
corresponds to its variance from $- N / 2$ to $0$ then back to $N / 2$.

Except the resemblance of their appearance, these two equations are quite 
different.  The first one is the quantized version of the second one and it is 
the evolution equation for quantum operators rather than classical variables.  
We can approximate the operators with their expectation values, with the 
lowest-order approximation corresponding the MFT and the second-order 
approximation corresponding to the equation given by  Ref.~\refcite{1}. 
Also two diagonal
terms of $-i U_0$ emerge in the quantized version, which
puts a phase factor to $J_x$ and $J_y$. So the phases are also important in 
the dynamics.  
The phases and particle numbers can be investigated under the mean field 
approximation~\cite{21}, where they exhibit oscillations in the phase space. 
In the angular momentum space, we can choose the eigenstates of $\hat{J}_z$ as the basis 
set $\left\{\left\vert j m \right\rangle \right\}$. The state  
of the system at time $t$ is given by 
$\left\vert \Psi \left( t \right) \right\rangle 
= exp \left\{- i H t \right\} \left\vert \Psi_0 \right\rangle$, 
where
$\left\vert \Psi_0 \right\rangle = \left\vert \frac{N}{2} \frac{-N}{2} \right\rangle $ is 
the initial state
with the left well
population.  The adiabatic transport  of the system then means the evolution from 
$\left\vert \Psi_0 \right\rangle$ to
$\left\vert \Psi \left( t  \right) \right\rangle 
=\left\vert \frac{N}{2} \frac{N}{2} \right\rangle$.

\section{Higher Order Interactions}

The above quantized version of Ref.~\refcite{2}  only includes the
nonlinear effects up to the on-site interaction.  It would be  interesting to 
investigate the effects of the neglected
interactions, which becomes useful when its behavior is investigated under a wide 
parameter regime. To do this, we add to the Hamiltonian
``Eq.~(\ref{hamiltonianBoson})'' the overlapping 
part
$ \phi_L^{\ast}\phi_R $
and introduce two new parameters $ U_t $ and 
$U_{tt}$ to describe their effects, where $ U_t $ captures interactions 
between the tunneling particle 
and the on-site particle and
$U_{tt}$ captures interactions between the tunneling particles themselves. 
We term the tunneling particle as 'tunnelier' for simplicity. 
Now the full Hamiltonian is given by 
\begin{eqnarray}
\mathcal{H} & = & \epsilon_L \hat{a}_L^{\dagger} \hat{a}_L +  \Omega \hat{a}_L^{\dagger} \hat{a}_R +
\frac{U_0}{2} \hat{a}_L^{\dagger} \hat{a}_L^{\dagger} \hat{a}_L
\hat{a}_L\nonumber\\
& &{}  + U_t \left(  \hat{a}_L^{\dagger} \hat{a}_L^{\dagger} \hat{a}_L
\hat{a}_R +  \hat{a}_L^{\dagger} \hat{a}_R^{\dagger} \hat{a}_L \hat{a}_L \right) 
 + \frac{U_{tt}}{2} \left(
\hat{a}_L^{\dagger} \hat{a}_L^{\dagger} \hat{a}_R \hat{a}_R + 2 \hat{a}_L^{\dagger} \hat{a}_R^{\dagger}
\hat{a}_R \hat{a}_L \right) \nonumber\\
                      &     &{} + \epsilon_R \hat{a}_R^{\dagger} \hat{a}_R  + \Omega \hat{a}_R^{\dagger}
                      \hat{a}_L \frac{U_0}{2} \hat{a}_R^{\dagger} \hat{a}_R^{\dagger} \hat{a}_R
\hat{a}_R \nonumber\\
& &{} + U_t \left( \hat{a}_R^{\dagger}
\hat{a}_R^{\dagger} \hat{a}_R \hat{a}_L + \hat{a}_R^{\dagger} \hat{a}_L^{\dagger} \hat{a}_R \hat{a}_R \right)
+ \frac{U_{tt}}{2} \left( \hat{a}_R^{\dagger} \hat{a}_R^{\dagger} \hat{a}_L \hat{a}_L + 2 \hat{a}_R^{\dagger}
  \hat{a}_L^{\dagger} \hat{a}_L \hat{a}_R \right),
\end{eqnarray}
with the newly introduced parameters given by
$U_t = g {\displaystyle \int d^3r\, \left\vert \psi_L \right\vert ^2 \left( \psi_L^{\ast} \psi_R \right)}
  = g {\displaystyle \int d^3r\, \left\vert \psi_R \right\vert ^2 \left( \psi_R^{\ast} \psi_L \right)}$ and
    $U_{tt} =g {\displaystyle \int d^3r\, \left( \psi_L^{\ast} \psi_R \right)^2} = g {\displaystyle \int
      d^3r\, \left( \psi_R^{\ast} \psi_L \right)^2} = g {\displaystyle \int d^3r\, \psi_L^{\ast} \psi_R
      \psi_R^{\ast} \psi_L}$.
For simplicity reasons, we have assumed $\psi_L^{\ast} \psi_R = \psi_R^{\ast} \psi_L$.
 In the Schwinger representation, this Hamiltonian becomes
\begin{eqnarray}\label{fullHamiltonian}
\mathcal{H} & = & \frac{E}{2} N + \epsilon \hat{J}_{z} + 2 \Omega \hat{J}_{x} + U_0 \left( \hat{J}_z^2 + \frac{N^2}{4} -
  \frac{N}{2} \right) \nonumber\\
&  & {} + 2 U_t \left( N - 1 \right) \hat{J}_x + \frac{U_{tt}}{2} \left( \hat{J}_- \hat{J}_- + \hat{J}_+
  \hat{J}_+ + 2 \left(  \hat{J}_+ \hat{J}_- + \hat{J}_- \hat{J}_+ \right) - 2 N \right)\nonumber\\
& = &\frac{E}{2} N + \epsilon \hat{J}_{z} + 2 \Omega \hat{J}_{x} + U_0 \left( \hat{J}_z^2 + \frac{N^2}{4} -
  \frac{N}{2} \right) \nonumber\\
&  & {} + 2 U_t \left( N - 1 \right) \hat{J}_x + U_{tt} \left(  \hat{J}_x^2 -
  \hat{J}_y^2 - 2 \hat{J}_z^2 +\frac{N^2}{2}\right)\nonumber\\ 
                      & = &\left( U_0 - 2 U_{tt} \right) \hat{J}_z^2 + \epsilon \hat{J}_z + 2 \left( \Omega +
                        U_t \left( N - 1 \right) \right) \hat{J}_x +  U_{tt} \left( \hat{J}_x^2 - \hat{J}_y^2 \right).
\end{eqnarray}
This representation illustrates the  interactions between the tunneliers, with $\hat{J}_- \hat{J}_-$ the
interaction between left tunneliers (particles tunneling from the right well to the left well),
$\hat{J}_+ \hat{J}_+$ the interaction between the right 
tunneliers (particles tunneling from the left well to the right well),
and $\hat{J}_+ \hat{J}_- + \hat{J}_-
 \hat{J}_+$ the interaction between the left tunnelier and the right tunnelier.
 We can see the modifications introduced to the
Hamiltonian by comparing it with ``Eq.~(\ref{simpleAngular})''.
The interaction between the tunnelier and the on-site particle adds a term $U_t \left( N - 1 \right)
$ to the original tunneling parameter $\Omega$,   the interaction between the left and the right tunneliers
adds a term $- 2 U_{tt}$ to the original nonlinear on-site interaction parameter $U_0$ and the interaction of
the left tunneliers  and that of the right tunneliers add another nonlinear terms $U_{tt} \left( \hat{J}_x^2
  - \hat{J}_y^2 \right)$ to the whole Hamiltonian.
The equation of motion of these operators is
\begin{equation} \label{fullDynamics}
\frac{d}{dt} \left( \begin{array}{c} \hat{J}_x\\
\hat{J}_y\\
\hat{J}_z \end{array} \right) = \left( \begin{array}{ccc} - i\left(U_0 - U_{tt}\right) & - \left( \epsilon +
  2\left(U_0 - U_{tt}\right) \hat{J}_z \right) & 0\\
\epsilon + 2\left(U_0 - 3 U_{tt}\right) \hat{J}_z & - i\left(U_0 - 3 U_{tt}\right) & - 2\left( \Omega + U_t
  \left(N - 1\right)\right)\\
0 &  2\left( \Omega + U_t
  \left(N - 1\right) + 2 U_{tt} \hat{J}_x\right) & - i 2 U_{tt} \end{array} \right) \left( \begin{array}{c} \hat{J}_x\\
\hat{J}_y\\
\hat{J}_z \end{array} \right).
\end{equation}

Compared with ``Eq.~(\ref{simpleDynamics})'', this equation is non-symmetric and there is even a phase term of
$\hat{J}_z$. By the definition of $\hat{J}_z$, there is no phase term as the phases carried by $\hat{a}_{L,
  R}$ is canceled out in $\hat{a}_L^{\dagger} \hat{a}_L$ and $\hat{a}_R^{\dagger} \hat{a}_R$. The origin of
this phase term of $\hat{J}_z$ manifests in the commutation relationship of $\hat{J}_x$ and $\hat{J}_y$, which
is introduced in deriving this equation of motion from the Hamiltonian ``Eq.~(\ref{fullHamiltonian})''.

In Ref.~\refcite{2}, the nonlinear parameter $U_0$ generates 
a rich adiabatic dynamics of the double-well BEC. Here the quantum corrections 
introduced by $U_{tt}$ and $U_t$ makes the dynamics even more complex. 
When $U_{tt}$ and $U_t$ are
small compared with $U_0$, the influence may not be  very 
explicit; but when they are large enough, the formulations above would lose sense as the
two-mode  assumption no longer holds true. Numerical simulations are  
needed to investigate their influence to the dynamics under various parameter
regimes. 

\section{Conclusion}

We have obtained quantum corrections of higher order interactions for the 
quantized dynamics of the double-well BEC and given a fully quantized 
expression of it under the two-mode approximation.  Two new parameters are 
obtained to express the influence of the interactions between the tunneling 
particle and the on-site particle and between the tunneling particles.  So the 
dynamics of the double-well BEC becomes more rich under this quantized 
version.  This allowes numerical simulation for a wide range of parameter 
regime and new phenomenon may be obtained.  

\section*{Acknowledgements}
Project supported by the National Natural Science Foundation of China (Grant
No.~11105086), the Natural Science Foundation of Shandong Province (Grant No.~BS2011DX029,
ZR2009AM026), the basic scientific research business expenses of the central university, and
the basic scientific research project of Qingdao(Grant No.~11-2-4-4-(6)-jch).

\end{document}